# A FRAMEWORK FOR PROCESS ASSESSMENT OF SOFTWARE PRODUCT LINE

**FAHEEM AHMED, University of Western Ontario**
Department of Electrical & Computer Engineering, London Ontario, Canada, N5A5B9,
Email: fahmed@engga.uwo.ca, Tel: 1-(519) 661-2111 Ext (81412), Fax: 1-519-850-2436

**LUIZ FERNANDO CAPRETZ, University of Western Ontario**
Department of Electrical & Computer Engineering, London Ontario, Canada, N5A5B9, Email: lcapretz@eng.uwo.ca,
Tel: 1-(519) 661-2111 Ext (85482), Fax: 1-519-850-2436

## ABSTRACT

Software product line has emerged as an attractive phenomenon within organizations dealing with software development process. It involves assembly of products from existing core assets, commonly known as components, and continuous growth in the core assets as production proceeds. Organizations trying to incorporate the concept of software product line to reduce development time and cost require certain rules to be followed for successful development and management, they also require a direct procedure to evaluate the current maturity level of the process. In this work certain rules for developing and managing a software product line are put forward. Additionally, a fuzzy logic based software product line process assessment tool (SPLPAT) has been designed and implemented on the basis of developed rules for software product line process assessment. SPLPAT can be used to assess the process maturity level of software product line, and it provides an opportunity to handle imprecision and uncertainty present in software process variables. Four case studies were conducted to validate the framework, and results show that SPLPAT provides a direct mechanism to evaluate current software product line process maturity level within an organization. The results of the developed software product line process assessment approach were compared with the existing CMM-level of the organization in order to evaluate the reliability of the presented approach and to find out how effectively an organization can execute software product line process when it has already achieved a certain CMM level.

## INTRODUCTION

The concept of software product line is based on development of identical systems having controlled variability among one another. The term "software product line" is widely used in North America whereas a similar concept but with different terminology like "product family" or "system family" is being used in Europe (Linden 2002). A software product line is a set of software-intensive systems sharing a common, managed set of features that satisfy the specific needs of a particular market segment or mission and that are developed from a common set of core assets in a prescribed way (Clements 2002). Software product line is gaining popularity over the time due to economics, but there has not been much research to establish appropriate rules as guidelines for software product line development or to come up with procedures to assess the maturity level of software product line process within an organization.

The aim of this research is to introduce a set of rules based on best-known practices of the software industry as well as to create a fuzzy logic-based framework and tool for process assessment of software product line within an organization. The focus of the software process assessment framework is to put forward a methodology for process assessment, particularly that of software product line. The correlation between CMM and the presented approach is beyond the scope of this work. The work presented in this paper does not propose an alternate methodology for CMM; rather it concentrates on developing a methodology for process assessment of software product line only. The consideration of CMM in this work addresses

## CONTRIBUTION

The research work presented in this paper makes a considerable contribution to information technology, particularly in the area of software product line, which is a relatively new methodology of software development. The major contributions are as follows:

- The software product line qualification rules presented in this paper aim at filling the research gap between concise guidelines and appropriate rules for developing and managing software product line.

- These rules assist an organization in developing an infrastructure for software product line, based on certain process requirements and on continuously monitoring the process at development and management levels.

- The software product line process assessment approach presented in this research work is novel; there had been no work done yet in this area, to the best of our knowledge.

- This work identifies the key process areas of software product line, ones that are used to perform the software process assessment within an organization.

- The knowledge-based software process assessment will help organizations to handle the imprecise and uncertain nature of software process data, and it provides them a more reliable assessment.

- The developed tool, SPLPAT can be used as a direct measurement approach for process assessment, particularly to software product line development and fills the research gap of process assessment approach to software product line.

This work is expected to be very interesting to those performing research in software process assessment practice and its improvements. The software community, particularly developers, managers and practitioners will benefit by understanding the software product line process in a concise and prescribed way. This research will help them to evaluate their current process maturity level, and this in turn will assist management's decision-making process in their efforts to improve the productivity of the development process.

the following:

- In order to evaluate the reliability of the proposed approach, we compared the software product line process assessment with the existing CMM levels achieved by the organizations under study. The fuzzy logic approach presented in this work transforms the software product line process variables into CMM levels as output. The purpose of this transformation is to investigate the extent of reliability of the proposed approach and compare it with an existing standardized approach like CMM.

- Another aspect of CMM involvement with this presented approach is to investigate the impact of already achieved CMM level on software product line process. The case studies presented in this paper are used to find out how effectively an organization can execute a software product line process when it has already achieved a higher CMM level.

- The imprecision and uncertainty present in software process data is investigated and the lesson learned from case studies presented in this work supports the concept that fuzzy logic use in software product line process assessment handles imprecision and uncertainty in a much better way compared to other standard methods; it also provides more reliable assessment.

## FUZZY LOGIC AND SOFTWARE PROCESS ASSESSMENT

The term fuzzy logic was introduced by Zadeh (1965) to handle situations where precise true/false values cannot be determined. Fuzzy logic is a form of algebra, one that deals with a range of values from "true" to "false" for the purpose of decision-making, using imprecise data. Zadeh (1992) elaborated to say that the purpose of fuzzy logic is to provide concepts and techniques that represent and derive knowledge that is imprecise, uncertain or lacking reliability. Software process is defined by certain activities performed at development and management levels. Maturity assessment of software process requires quantitative data about how effectively those activities are being performed. For example, if we consider an activity like "requirements engineering" performed during software process then we cannot apply traditional "yes/no" answers or "zero/one" logic to it, because they indicate only that requirements engineering is done or not.

Figure 1 represents boolean logic for assessment of maturity for requirements engineering; it handles only two states: either the activity is performed or not- there is no intermediate stage present, and the fact is that requirements engineering might be performed partially but not completely. In order to handle this imprecision and uncertainty, Figure 2 illustrates the use of fuzzy logic to represent quantitative assessment of requirements engineering activity over the range of values. The membership function in the domain of 0 to 1 precisely defines how much the activity is performed, for example 0.3 indicates slightly performed and 0.95 illustrates definitely performed.

Research has been done in order to address the imprecision and uncertainty present in software process variables. Büyüközkan, Kahraman and Ruan (2004) proposed a methodology to improve the quality of decision-making in software development project under uncertain conditions. The proposed methodology is based on fuzzy analytic hierarchy process modeling to deal with uncertainty and vagueness arising from subjective perception and experience of humans in the decision process. Using a fuzzy logic based approach; Cimpan and Oquendo (2000) monitored software processes and concluded that fuzzy logic offers significant advantages over other approaches due to its ability to naturally represent uncertain and imprecise information. Sun Sup, Sung Deok, and Yong Rae (2002) proposed a fuzzy logic based software quality prediction model to analyze the software process data for the purpose of software process improvement. Ziv and Richardson (1997) presented the concept of "maxim of uncertainty in software engineering" (MUSE), which states that uncertainty is abundant and inevitable in software development. Xiaoqing, Kane and Bambroo (2003) put forward an intelligent software early warning system based on fuzzy logic, one that assesses risks associated with poor quality in software development. The proposed approach aims at handling incomplete, inaccurate, and imprecise information, and resolves conflicts in an uncertain environment.

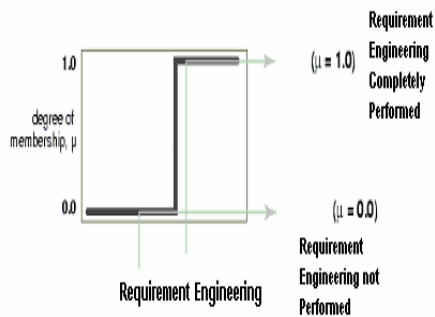

**Figure 1. Boolean Logic for Requirement Engineering Activity Assessment**

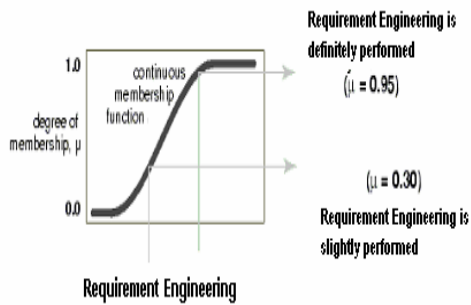

**Figure 2. Fuzzy Logic for Requirement Engineering Activity Assessment**

## SOFTWARE PRODUCT LINE RULES

The concept of software product line has become an attractive phenomenon within organizations dealing with software development. Organizations attempting to incorporate this concept require certain rules to be followed for effective development and management. There is a need to define and summarize all the necessary guidelines and principles for software product line development and management activities as set of rules so that they should be carefully followed for successful outcomes. The rules are categorized as core asset development rules, product development rules, and management rules, and cover the three essential activities of software product line development and management. The general structure of the rules consists in a "statement portion" and a "discussion portion".

**Statement:** defines the rule

**Discussion:** elaborates the rule, with possible implications of not following it.

## CORE ASSET DEVELOPMENT RULES

Core assets are developed to create products in a software product line. The most vital core asset in the repository is architecture, because all subsequent products must share it. These rules describe the general principles in developing and managing a core asset repository for further reuse of core assets during software product development activity. These rules elaborate the qualification criteria of the components present in the core asset repository. They highlight the importance of managing a version control and utilization history of the core asset in order to track all the entities.

### Core Asset Development Rule # 1

**Statement:** "All the core assets within a software product line repository and resulting products must be consistent with the scope of the software product line."

**Discussion:** DeBaud and Schmid (1999) stressed that it is important to define the proper scope of the software product lines, as it is necessary for the strategic development of product lines. Kishi, Noda and Katayama (2002) found that the principal use of product line scope is to define the product line and the products that comprise the product line. Once the scope of software product line is defined it is necessary that the entire core assets be consistent with the scope of product line. The aim is to develop core asset for the product line within the scope of the product line so that they can be utilized while developing products. The products developed should fall within the scope of the product line as well, so that product line requirements are met.

## Core Asset Development Rule # 2

**Statement:** "Every component present in the core asset repository must define the variability mechanism to tailor it for effective utilization."

**Discussion:** According to Czarnecki and Eisenecker (2000), instead of developing and deploying a "fixed" one-of-a-kind system, it is now common to develop a family of systems whose members differ with respect to functionality or technical facilities offered. Fitting the component into the product without tailoring it is the easiest task, but often there arises a need to make certain changes in the components to meet the requirements for a particular product. Every component present in the core asset must clearly define the variability mechanism to be used in order to tailor it for reuse. A separate document must be attached with each component, one that elaborates this activity. A careful strategy should be adopted to accommodate variability among components so that changes will remain within the scope of the product line.

## Core Asset Development Rule # 3

**Statement:** "Update core asset repository constantly by adding new asset as product lines progress."

**Discussion:** If we use a proactive approach to develop software product line, then initially all the core assets are identified, and, as we progress further, products resulting from the product line tend to develop new core assets which must be added to the repository so that they can be reused for later products. If we use an active approach to develop product lines, we start developing products and core asset generated during the development process that constitutes the core asset repository. So whatever approach we adopt to develop software product line, the core asset repository should be dynamic and should continue increasing its size with the addition of components. The information about the updates of core asset repository must be clearly and regularly communicated to the developers.

## Core Asset Development Rule # 4

**Statement:** "All the COTS present in or added to core asset repository must satisfy the cost benefit ratio for the organization."

**Discussion:** COTS are software components developed by vendors to provide specific functionality; they can be used as a part of a product and are readily available in the market. Hall and Naff (2001) emphasized that use of COTS elements can result in reduced development cost, development and integration risk, and development time. Voas (1998) pointed out that developing software with as much COTS functionality as possible saves developers from reinventing the wheel, but at the same time it is required that the benefit-to-cost ratio satisfy the organizational goal; otherwise it will considerably increase the overall cost of the product. To achieve the target of product development within budget, all the COTS added to the core asset must meet the cost-to-benefit ratio criteria for the organization.

## Core Asset Development Rule # 5

**Statement:** "A version control management system should keep track of the core asset development and utilization history."

**Discussion:** Ambriola, Bendix and Ciancarini (1990) confirmed that the goals of version controls are to facilitate the efficient retrieval and storage of several versions of the same components and to enforce restrictions on the evolution of a component so that such an evolution is observable and controllable. The core assets in the repository are to be used in various products and their versions. It is necessary to track the history of the utilization of individual core asset in different products. This history should clearly describe the functionalities used, interface requirements, and any modifications to accommodate the core asset to a new product. If any considerable modification is made, it should be termed a new version of the same core asset and ultimately added to the core asset repository with an associated definition of its parent.

# PRODUCT DEVELOPMENT RULES

Products developed during software product line are those viable entities that can be utilized within an organization for specific purposes or that can be placed in the market to capture a business segment. These rules describe the qualification criteria for the products developed during software product line activity based on the core asset present in the core asset repository. These rules further elaborate the scope of the products developed and characteristics of the software product line. These rules define the extent of the difference among various products developed. Since software product line is capable of producing a number of products, the rules indicate that at least more than one product should be developed.

## Product Development Rule # 1

**Statement:** "All the products within the software product line must share a common architecture."

**Discussion:** Gomma and Farrukh (1999) described software architecture as the overall organization of a software system in terms of its constituent elements, including computational units and their interrelationships. Bass, Clements and Kazman (1998) stated that software architectures for product lines require open architectures and reuse design approach. The purpose of the software product line is not just reuse; it targets the effective delivery of shared architecture products to meet the market demands from business interests. It is necessary that all the products resulting from product line share a common architecture, the only criterion for them to be a part of the family of products. This commonality among the products, despite other differences or variability, creates a product population.

## Product Development Rule # 2

**Statement:** "A variation among products should remain within the scope of software product line."

**Discussion:** Robak and Pieczynski (2003) observed that possible features of software product line members vary according to the needs of particular market segments or purposes. The products from the software product line may vary from one another in quality, reliability, functionality, performance, etc, but, as they share the common architecture, the variation should not be so great that they are no longer contained within the scope of the product line. Those variations must be handled systematically to accommodate changes in various versions of the product. Variability control mechanism should allow changes and new features to be implemented in the resulting products, thereby keeping them within the scope of the product line.

## Product Development Rule # 3

**Statement:** "Every product released from product line should be a valid business case for the organization."

**Discussion:** The business cases define the marketing strategy of the organization; they explore the market for profitable business. Boeckle (2002) pointed out that goals of software product line are elaborated in business cases, and they promote the product line idea. John and Schmid (2001) concluded that the decision of launching software product line within an organization is based heavily on deciding whether the product line development will produce more benefits than its implementation cost. Every organization identifies potential business cases in order to capture the market. It is necessary that each product released from the software product line be a valid business case for the organization so that the organization can ultimately achieve its financial goal along with the justification of the product line itself.

## Product Development Rule # 4

**Statement:** "Software product line must capable of producing a considerable number of products, at least more than one."

**Discussion:** The main aim driving software product line is the development of a stream of products from core asset. If the product line is aimed to produce only one single product, then the activity can be regarded as "just a component-based development", not a software product line. Therefore the scope and structure of the product line should be to develop all those products that meet the business case criteria for the organization. Bandinelli (2001) investigated the idea that adoption of product line engineering implies the creation of a domain infrastructure, including architectures, components, training etc., all of which generally requires a significant up-front investment. Software product line must produce a considerable number of products to gain profits and capture market segment in order to justify the up-front cost of adoption.

### Product Development Rule # 5

**Statement:** "Every product released from the software product line must meet the qualification criteria of the organization."

**Discussion:** Every organization defines its parameters for the qualification of a product along with the standard acceptance criteria. A product is feasible only if it meets the qualification criteria as defined earlier in its development. Therefore the qualification criteria of the software product line must be clearly defined so that all the products resulting from the development of software product line meet those criteria. A comprehensive quality assurance plan must operate at all the organizational levels and should encompass all the possible stages of development from component to product, all the way up to product line.

## MANAGEMENT RULES

Management rules describe the essential management activities that must be followed and implemented for effective utilization of the software product line concept. Also, they integrate the associated processes that are required for software product development and management. The management rules address both the organizational and technical management domains. They describe some of the essential technical processes that have to be managed and introduced at all levels of the organization.

### Management Rule # 1

**Statement:** "A multi dimensional configuration management approach should handle the configuration management issues present in the software product line."

**Discussion:** Zhang, Mei and Zhu (2001) reported that configuration management system manages the artifacts produced in the development process, controls changes to the software and its components, keeps track of evolution, and thus assists the development process. Configuration management issues are imperative to address in software product lines, as they deal with a number of resultant products having different version and release numbers as well as numerous core asset with different versions. Therefore a multi-dimensional approach of configuration management should be adopted to cope with the issues. Such an environment may be defined as configuration management of the configuration management system. In this approach, separate configuration management systems are applied to product and core asset, and, on top of those two configuration managements, another configuration management handles the coordinated issues of both.

### Management Rule # 2

**Statement:** "A comprehensive description and analysis of domain should be performed for which the software product line is to be developed."

**Discussion:** John, Muthig, Sody and Tolzmann (2002) considered that the goals of a domain-analysis approach is to identify and document requirements of a set of systems in the same application domain in order to make development and maintenance activities more efficient. Comer (1990) found that domain analysis is the systems engineering of a family of systems in an application domain through development and application of reusable asset. Comer (1990) stressed that domain analysis entails developing a complete and rigorous domain model and associated generic architecture as a precursor to developing a set of reusable components for repeated application in developing systems in the domain. The domain analysis for software product line will support the development as well as reusing the core asset in development.

### Management Rule # 3

**Statement:** "The return on investment (ROI) of the software product line must meet the organizational financial goal."

**Discussion:** The construction of software product line will be beneficial in terms of finance only if the organizational ROI meets the outcome of software product line. The investment in the product line must justify itself. It is generally agreed that the return on software product line is heavily based on the resultant products and gradually increases as the number of products to be delivered to the market increases. Buckle, Clements, McGregor, Muthig and Schmid (2004) claimed that product line engineering could improve ROI from a set of products. The ROI of product line project is heavily dependent on which kind of approach has been used to institutionalize software product line. For example, in the proactive approach the upfront cost is lower because first a core asset repository is first established and then products are built from those assets.

**Management Rule # 4**

**Statement:** "Requirements of the software product line must be clearly defined, analyzed, specified, verified and managed."

**Discussion:** Gause and Weinberg (1989) reported that in the first stage of software project, usually requirements, elicitation, information and knowledge of the system under construction are acquired. Kruchten (1998) observed requirements management process as a set of three activities: eliciting, organizing, and documenting the system's required functionality and constraints; evaluating changes to these requirements and assessing their impact; tracking and documenting tradeoffs and decisions. John and Dorr (2003) pointed out that especially when developing more than one product, requirements elicitation is a complex task; in depth knowledge of the problem domain often is a prerequisite for a successful product family. If we perform good requirement management for the software product line, it will assist in understanding the scope and boundaries of the products to be developed, an understanding which ultimately aids in identifying core asset for the software product line.

**Management Rule # 5**

**Statement:** "Requirements of the software product line must define the fundamental products and their features within the product line."

**Discussion:** Bertolino, Fantechi, Gnesi, Lami and Maccari (2002) concluded that product family requirements could, in general, be considered as composed of a constant and a variable part. The constant part includes all those requirements dealing with features or functionalities common to all the products belonging to the family and which need not be modified. The variable part represents those functionalities that can be changed to differentiate one product from another. Software product line requirements define the features of the products in the product line. The engineering requirements of software product line must yield the features of fundamental product. It should describe the core functionality which the products are supposed to provide, the properties they must exhibit, and the associated constraints and quality parameters.

**Management Rule # 6**

**Statement**: "Organizational structure must support the software product line concepts and principles."

**Discussion:** Software product line approach is somewhat different from traditional software development approach. It requires considerable visualization, management enforcement, communications, discussions and elaborations of what must be done and what can be done. Therefore, organizational structures must support the concepts and principles of software product line. Boeckle (2002) found that transforming an organization to create products as members of a product family required installing corresponding processes, organization structures, and methods. The integral part of organization structure is design teams; small groups of people whose capabilities complement one another and which are formed for a common goal. There should be a clear definition of the team and its associated tasks and duties.

**Management Rule # 7**

**Statement:** "All the three essential activities of software product line development must be performed iteratively."

**Discussion:** Miller, Paradis and Whalen (1991) found that iterative process minimizes both risks and cost by combining both the well-structured management techniques of the waterfall process and the early validation techniques of the evolutionary model. Fujii and Kambayashi (2002) concluded that iterative development process proceeds with multiple iterations that typically consist of one cycle of development activities such as analysis, design, implementation, and testing. The essential activities, core asset development, management and product development, performed during product line development are linked and are highly iterative. Core asset are used to develop new products, and there is a continuing possibility of accumulating large numbers of core asset either as an outcome of new product development or COTS. The management takes its inputs from core asset and the development phase and continuously gives feedback to both. The iterative development approach will allow interaction among these three essential activities to support them.

## SOFTWARE PRODUCT LINE RULES: REMARKS

The rules presented in this section cover some of the possible aspects of software product line process. They can be further enhanced to cover other aspects of software product line development and management by adding more rules. Each rule has the tendency to be further divided into smaller sub-rules based on the activities performed in a software product line process. The further subdivision into smaller domains would identify software product line process at singleton level, which would further help in understanding the whole product line process. The further division of the rules into smaller parts is beyond the scope of this work.

## DESIGN AND IMPLEMENTATION OF SOFTWARE PRODUCT LINE PROCESS ASSESSMENT TOOL (SPLPAT)

A fuzzy logic based tool to measure performance of the process of software product line is designed and developed on the basis of rules for developing and managing software product line. Since software product line has three essential activities, i.e., core asset development, product development, and management, therefore the three-dimensional approach for process assessment of software product line as illustrated in Figure 3, illustrates that:

- Process assessment of an individual activity like product development, core asset development, and management are measured by using a fuzzy logic system and by applying software product line rules.

- Software product line process assessment is measured by applying the process assessment of individual activities like product development, core asset development and management to the fuzzy logic system.

Every fuzzy logic system requires certain inputs to process; therefore in order to take inputs in the form of rules for developing and managing software product line to the designed fuzzy logic system for process assessment, the following questions (Table 1) were asked. The crisp input to the fuzzy logic system depends on the values entered for each question in the range of 0 to 50. The value 0 reflects the lowest, whereas 50 the highest rating of the specified activity of the software product line process.

## ARCHITECTURE OF THE TWO-VARIABLE FUZZY LOGIC SYSTEM

Figure 4 illustrates a two-variable fuzzy logic system designed for software product line process assessment. It takes two variables as an input, and they can be any combination of two questions presented in Table 1. These two variables perform a fuzzification process which converts the crisp input into a fuzzy membership mapping that is applied to the inference engine, which in turn interacts with rule base to select the applicable rules based on the input variable values. The fuzzy output is then defuzzified to retrieve a crisp output. The rule base contains nine rules designed for software product line process assessment, which depicts an output based on the values of the input.

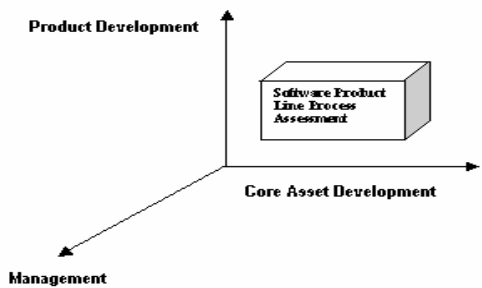

**Figure 3. Three Dimensional Process Assessment of Software Product Line**

The two variable approach of fuzzy logic is based on associative property of fuzzy sets. Since the questions presented in Table 1 can be further increased to accommodate other possible aspects of software product line, therefore this design choice can easily accommodate further expansion in input to the system. The input and output variables are represented by trapezoid function. The trapezoid function retains highest membership value of 1 up to a required interval. Using min operator carries out fuzzy implication, and using max-min operator performs composition. Centroid method is selected for defuzzyfication process. In centroid method, the crisp value of the output variable is computed by finding the variable value at the center of gravity of the membership function.

The questions presented in Table 1 are based on software product line rules; they cover some of the possible aspects of software product line process. One can add some more questions to cover other aspects of software product line development and management. Furthermore each question has the tendency to be further divided into smaller sub-level questions. The further sub-division into smaller questions would increase the precision of the input values, thereby avoiding possible biases associated with human judgment.

Figure 5 illustrates the basic architecture of the fuzzy logic based process assessment tool for software product line. Figure 5 illustrates that the questions presented in Table 1 are grouped together on the basis of activities like core asset development, product development, and management, and are applied to fuzzy logic system to get the process assessment of the three essential activities of software product line, which are:

- **Core Asset Development Assessment:** describes the maturity level of core asset development activity.

- **Product Development Process Assessment:** reflects the maturity level of product development activity.

- **Management Process Assessment:** shows the maturity level of technical and organizational management activity within an organization.

### Software Process Input Variable A Fuzzy Representation

The term crisp value in fuzzy logic system is used to represent any precise numerical value such as 2, –3, or 7.34 etc. The crisp input to the system is selected to fall in the range of 0 to 50. The crisp input values are divided into three linguistic categories, i.e., "yes", "no" and "partial", in the range of 0 to 50:

- **Yes:** means that the activity is completely performed and is represented in the range of 33.0 to 50.0.

- **Partial:** means that the activity is performed but not completely, and is represented in the range of 16.5 to 38.0.

- **No:** means that the activity is not performed and it is represented in the range of 0 to 21.5.

**Table 1. Software Product Line Process Assessment Input Question**

| | |
|---|---|
| **Core Asset Development Input Questions** | |
| Question 1. | Are all of the core assets within the software product line repository and the resultant products consistent with the scope of software product line? |
| Question 2. | Do all the components present in the core asset repository define the variability mechanism to tailor them for effective utilization? |
| Question 3. | Do all the COTS present or added into core asset repository satisfy the cost benefit ratio for the organization? |
| Question 4. | Is the core asset repository updated constantly by adding new asset as the product line progresses? |
| Question 5. | Does a version control management system keep track of the core asset development and utilization history? |
| **Product Development Input Questions** | |
| Question 6. | Do all the products within the software product line share a common architecture? |
| Question 7. | Does the variation among products remain within the scope of software product line? |
| Question 8. | Is every product released from the product line a valid business case for the organization? |
| Question 9. | Does the software product line produce a considerable number of products, or at least more than one? |
| Question 10. | Does every product released from the software product line meet the qualification criteria of the organization? |
| **Management Input Questions** | |
| Question 11. | Is any configuration management system used to address the configuration management issues present in the software product line? |
| Question 12. | Is a comprehensive description and analysis of domain performed for the software product line? |
| Question 13. | Does the ROI (Return on Investment) of the software product line meet the organization's financial goal? |
| Question 14. | Are the requirements of the software product line clearly defined, analyzed, specified, verified and managed? |
| Question 15. | Does the requirement of the software product line define the fundamental products and their features within the product line? |
| Question 16. | Does the organizational structure support the software product line concepts and principles? |
| Question 17. | Are the essential activities of software product line development performed iteratively? |

A trapezoid function is used to represent the mapping between fuzzy membership in the range of 0 to 1, and crisp input values in the range of 0 to 50. The Equation-I represents the mathematical model of the trapezoid function. The values of the variables a, b, c and d, in the equation, define the shape of the trapezoid. The graphical representation of trapezoid function along with variables a, b, c, and d, is represented in Figure 6, which illustrates that the choice of variables a, b, c and d determine the shape of the trapezoid. Table 2 shows the distribution of linguistic variables yes, no and partial in the range of 0 to 50, and describes the values for variables a, b, c and d, in Equation-I for a mapping between linguistic variable and fuzzy membership values. Figure 7 illustrates the distribution of input linguistic variables, yes, no and partial in the range of 0 to 50, and fuzzy membership mapping in the range of 0 to 1.

The distribution of values from 0 to 50 among the linguistic variables no, partial and yes is based on equality in the range to observe maximum membership of 1 and equal overlapping regions among them. The tolerance is ± 0.5. The trapezoid values "b" and "c" presented in Table 2 illustrates that all three linguistic variables no, partial and yes retain a fuzzy membership of 1 at a range of 11.5 ± 0.5. The overlapping regions of 5.0 among the linguistic variables are defined by trapezoid values "a" and "d" presented in Table 2.

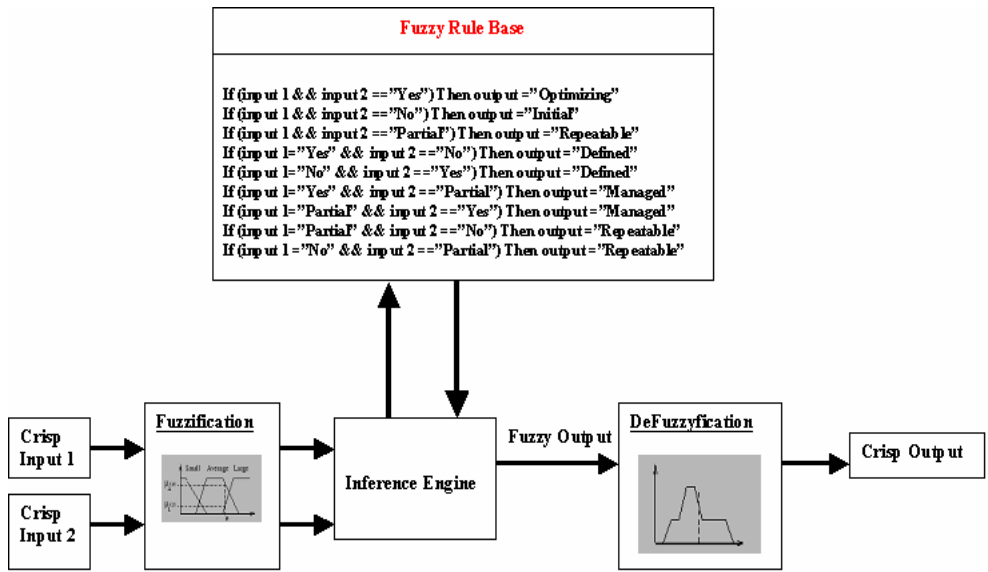

**Figure 4. Two Variable Fuzzy Logic System for Software Product Line**

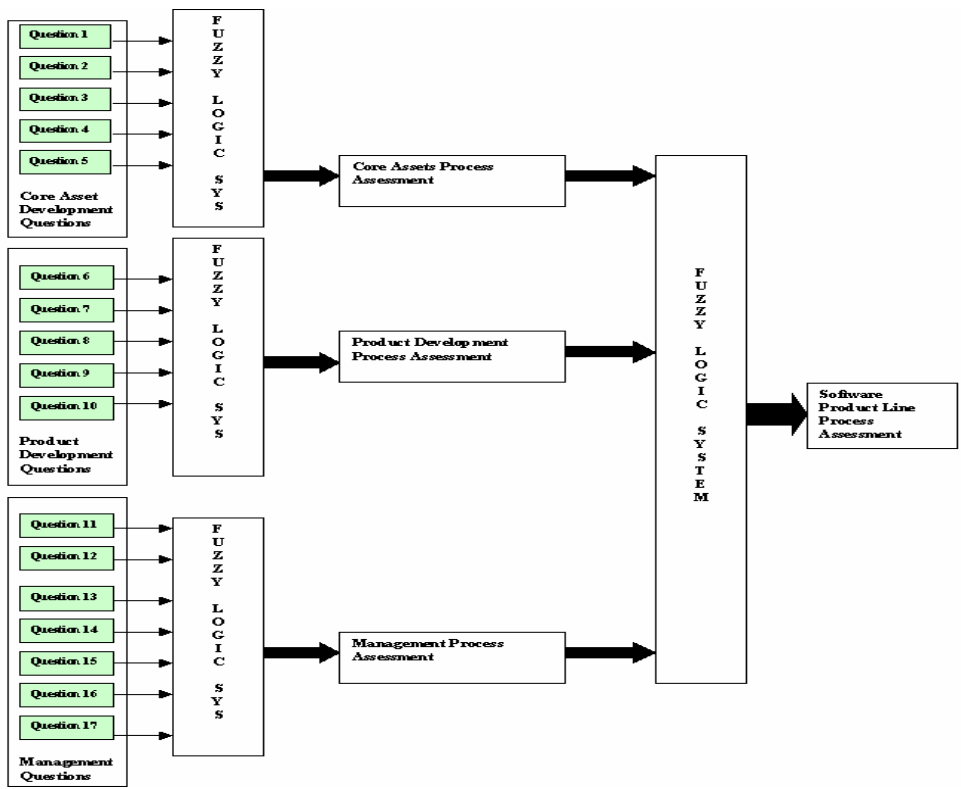

**Figure 5. Architectural View of SPLPAT**

$$f(x;a,b,c,d) = \begin{cases} 0 & \text{for } x < a \\ \dfrac{x-a}{b-a} & \text{for } a \le x < b \\ 1 & \text{for } b \le x < c \\ \dfrac{d-x}{d-c} & \text{for } c \le x < d \\ 0 & \text{for } d \le x \end{cases}$$

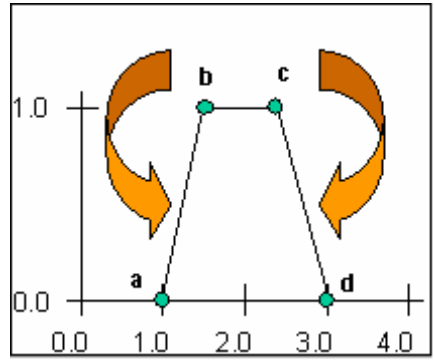

**Equation-I**                    **Figure 6. Trapezoid Function**

**Table 2. Input Values Linguistic, Crisp and Fuzzy Membership**

| Linguistic Value | Crisp Value Range | Trapezoid Function Variable Values For Input Fuzzy Membership Mapping | | | |
|---|---|---|---|---|---|
| | | **a** | **b** | **c** | **d** |
| Yes | 33.0 to 50 | 33.0 | 38.0 | 50.0 | 50.0 |
| No | 0 to 21.5 | 0.0 | 5.0 | 16.5 | 21.5 |
| Partial | 16.5 to 38.0 | 16.5 | 21.5 | 33.0 | 38.0 |

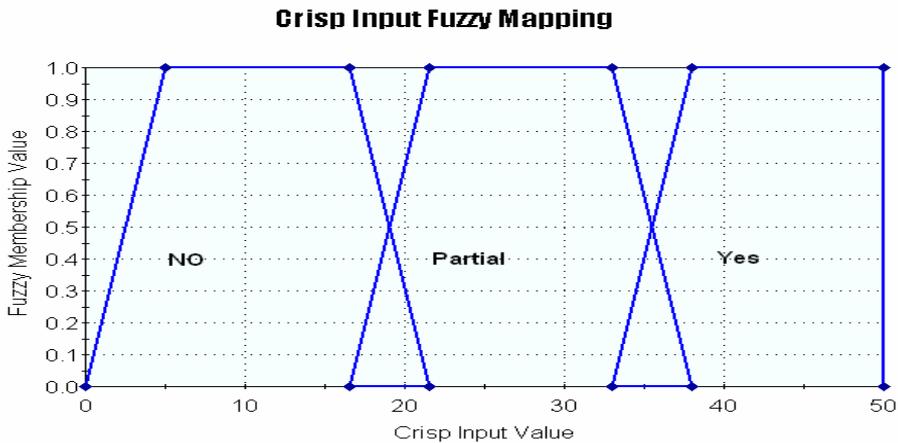

**Figure 7. SPLPAT Crisp Input -Fuzzy Membership Mapping**

## Software Process Output Variable A Fuzzy Representation

The crisp output of the system is selected to fall in the range of 0 to 50. The crisp output values are divided into five linguistic categories, i.e., initial, repeatable, defined, managed and optimizing in the range of 0 to 50, similar to the CMM approach, and described as following:

- **Initial:** defined in the interval of 0.0 to 15.0

- **Repeatable:** defined in the interval of 10.0 to 25.0

- **Defined:** defined in the interval of 20.0 to 35.0

- **Managed:** defined in the interval of 30.0 to 45.0

- **Optimizing:** defined in the interval of 40.0 to 50.0

A trapezoid function is used to represent the mapping between fuzzy membership in the range of 0 to 1, and crisp output values in the range of 0 to 50. Table 3 illustrates the distribution of linguistic output variables, initial, repeatable, defined, managed and optimizing in the range of 0 to 50 and shows the values for variables a, b, c and d, in Equation-I for a mapping between linguistic variable and fuzzy membership values. Figure 8 illustrates the distribution of output linguistic variables in the range of 0 to 50 and fuzzy membership mapping in the range of 0 to 1.

The distribution and choice of values from 0 to 50 among the linguistic variables initial, repeatable, defined, managed and optimizing is based on equality in the range to observe maximum membership of 1 and equal overlapping regions among them. The trapezoid values "b" and "c" presented in Table 3 illustrates that all the five linguistic variables, initial, repeatable, defined, managed and optimizing retain a fuzzy membership of 1 at a range of 5.0. The overlapping regions of 5.0 among the linguistic variables are defined by trapezoid values "a" and "d" presented in Table 3.

**Table 3. Output Values Linguistic, Crisp and Fuzzy Membership**

| Linguistic Value | Crisp Value Range | Trapezoid Function Variable Values For Output Fuzzy Membership Mapping | | | |
|---|---|---|---|---|---|
| | | **a** | **b** | **c** | **d** |
| Initial | 0.0 to 15.0 | 0.0 | 5.0 | 10.0 | 15.0 |
| Repeatable | 10.0 to 25.0 | 10.0 | 15.0 | 20.0 | 25.0 |
| Defined | 20.0 to 35.0 | 20.0 | 25.0 | 30.0 | 35.0 |
| Managed | 30.0 to 45.0 | 30.0 | 35.0 | 40.0 | 45.0 |
| Optimizing | 40.0 to 50.0 | 40.0 | 45.0 | 50.0 | 50.0 |

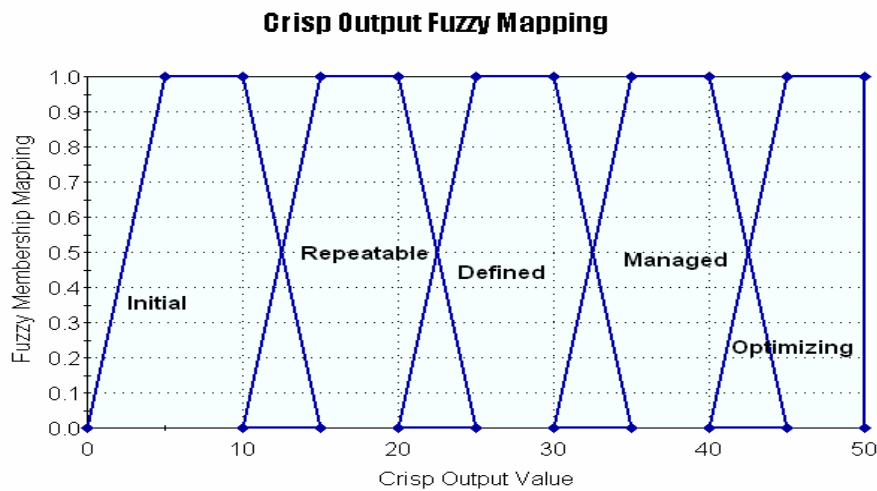

**Figure 8. SPLPAT Output Crisp-Fuzzy Membership Mapping**

**Fuzzy Logic Rules of Software Product Line Process Assessment**

The fuzzy knowledge rule base is created to contain fuzzy logic rules for fuzzy reasoning, particularly for software product line process assessment tool. The rules generally define a combination of input crisp pattern and respective output. On the basis of combination of input, appropriate output mapping is defined in the fuzzy logic rules. The variables defined as input 1 and input 2 can be any combination of questions presented in Table 1. There are nine rules for software product line process assessment tool, as listed below:

If (input 1 && input 2 =="Yes") Then output ="Optimizing"

If (input 1 && input 2 =="No") Then output ="Initial"

If (input 1 && input 2 =="Partial") Then output ="Repeatable"

If (input 1="Yes" && input 2 =="No") Then output ="Defined"

If (input 1="No" && input 2 =="Yes") Then output ="Defined"

If (input 1="Yes" && input 2 =="Partial") Then output ="Managed"

If (input 1="Partial" && input 2 =="Yes") Then output ="Managed"

If (input 1="Partial" && input 2 =="No") Then output ="Repeatable"

If (input 1 ="No" && input 2 =="Partial") Then output ="Repeatable"

Figure 9 illustrates the internal processing sequence of software product line process assessment tool, in which a combination of two questions from Table 1 are placed at the input of two-variable fuzzy logic systems described in Figure 4. The intermediate outputs are collected and passed to the two-variable fuzzy logic system at the next level and this procedure keeps on moving until we collect the individual software product line activity like core asset development, management, and product development assessment, which are later applied to next stage fuzzy logic system to eventually get the overall software product line process assessment.

## CASE STUDIES & CRITICAL DISCUSSION

Four case studies were conducted in order to validate the results achieved from the software product line process assessment tool. The input questionnaire (shown in Table 1) was distributed to a number of organizations to obtain actual data about current process status within the organization, as shown in Table 4. Some large and well-known organizations extensively involved in software development, under an agreement to keep the name of organization confidential, provided us with actual data. For experimental purposes the organizations are code named as "A", "B", etc. Organizations were asked to inform us of the actual CMM-level that they had achieved. The results observed from software product line process assessment tool are compared with the CMM-level achieved by the organization.

## DISCUSSION OF CASE STUDY –I

The CMM-level of organization "A" is level 2, i.e., "repeatable", and software product line process assessment tool has also evaluated it as level 2.The results in Table 5 indicate that core asset development activity is performed at a maturity level of 3 to 4, meaning that level 3 has been achieved and level 4 is close to achieve. Product development activity is performed at a much higher maturity of level 3. The management activity has a maturity level of 1, which lowers the overall process assessment to level 2, i.e., "repeatable".

Figure 10 describes the processing sequence and intermediate results at each of the stages during software product line process assessment of case study -I. The main conclusion of case study -I indicates that organization "A" can improve the overall software product line process by concentrating on management activity. A considerable improvement in the management process is required to improve the overall software product line process in organization "A". Figure 11 and Figure12 are input and output screens of SPLPAT.

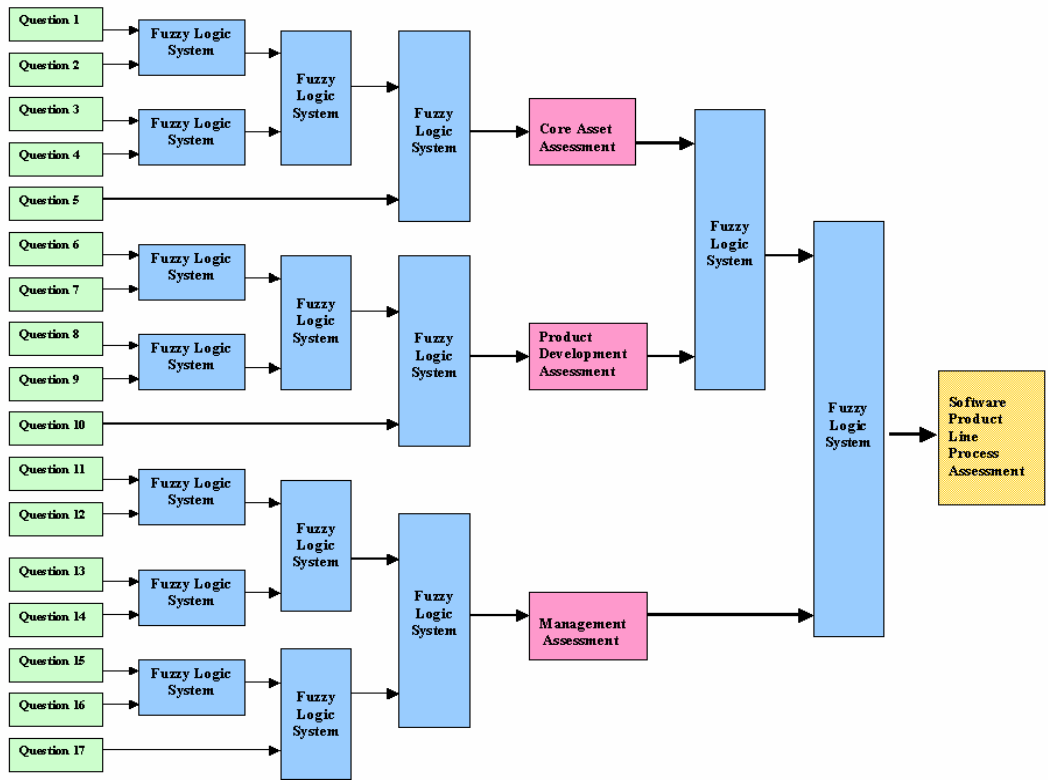

**Figure 9. Internal Processing Sequence of Software Product Line Process Assessment Tool**

**Table 4. Software Product Line Process Data of Case Studies**

| Rule-Input # | Organization  & Case Study # | | | |
|---|---|---|---|---|
| | 'A' | 'B' | 'C' | 'D' |
| | Case Study -1 | Case Study -2 | Case Study -3 | Case Study -4 |
| 1 | 35 | 40 | 32.5 | 40 |
| 2 | 40 | 40 | 27.5 | 30 |
| 3 | 25 | 15 | 30 | 35 |
| 4 | 35 | 30 | 37.5 | 30 |
| 5 | 25 | 50 | 40 | 20 |
| 6 | 40 | 15 | 37.5 | 40 |
| 7 | 10 | 15 | 32.5 | 35 |
| 8 | 5 | 30 | 30 | 35 |
| 9 | 50 | 50 | 35 | 30 |
| 10 | 45 | 40 | 37.5 | 30 |
| 11 | 30 | 50 | 32.5 | 25 |
| 12 | 10 | 40 | 35 | 20 |
| 13 | 15 | 40 | 30 | 30 |
| 14 | 20 | 30 | 35 | 35 |
| 15 | 30 | 40 | 32.5 | 35 |
| 16 | 35 | 45 | 30 | 35 |
| 17 | 7 | 25 | 37.5 | 35 |

**Table 5. Results of Software Product Line Process Assessment of Case Study-I**

| Activity | Result | Linguistic Output | CMM Level |
|----------|--------|-------------------|-----------|
| Core Asset Development Assessment | 34.84 | Defined to Managed | 3 to 4 |
| Product Development Process Assessment | 29.27 | Defined | 3 |
| Management Process Assessment | 8.64 | Initial | 1 |
| Software Product Line Process Assessment | 17.5 | Repeatable | 2 |

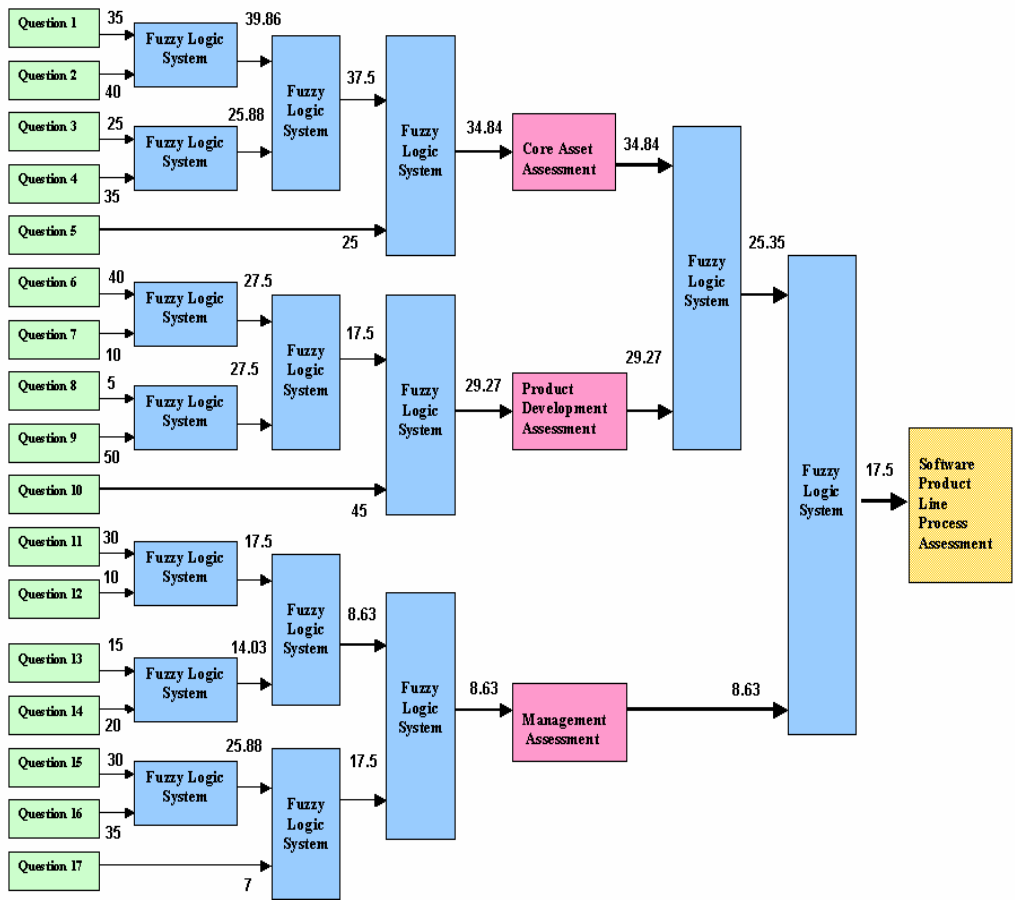

**Figure 10. Processing Sequence and intermediate Results for CASE Study -I**

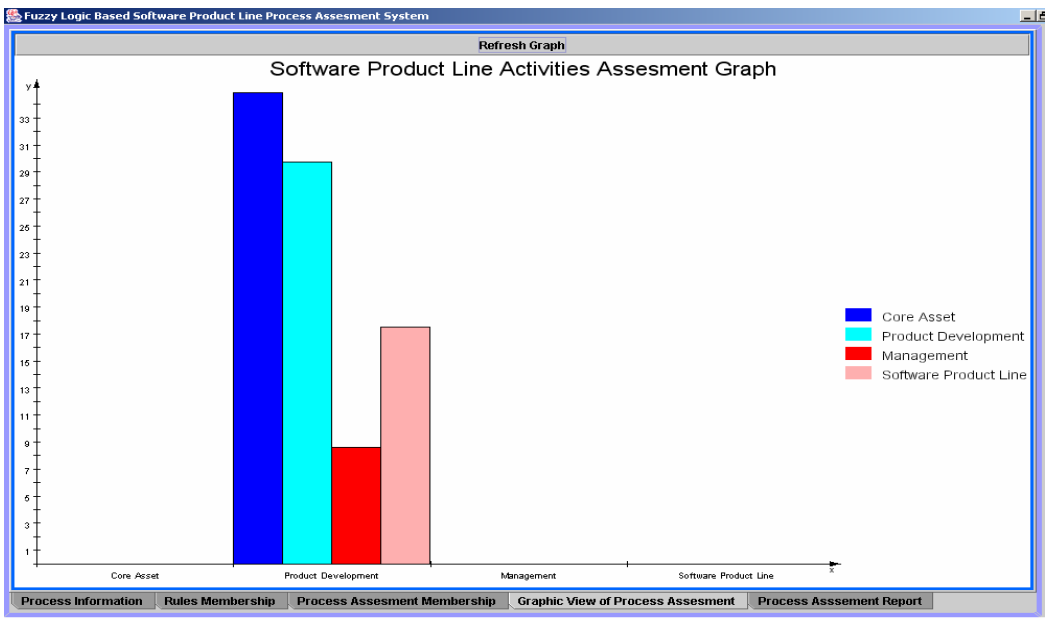

**Figure 11. Data Entry Input Screen of SPLPAT**

**Figure 12. Graphical Output Screen of SPLPAT**

## DISCUSSION OF CASE STUDY – II

The CMM-level of organization "B" is level 5, i.e., "optimizing" and the software product line process assessment tool has also evaluated it as level 5.The results in Table 6 show that core asset development activity is performed at a higher maturity level of 4. The product development activity is also performed at a very high maturity of level 4. The management activity is performed at a maturity level of 4 to 5, meaning they have achieved level 4 and now are very close to level 5. The conclusion of the case study –II highlights that the organization "B" has achieved a CMM-level 5.

## DISCUSSION OF CASE STUDY – III

The CMM-level of organization "C" is level 3, i.e., "defined", and the software product line process assessment tool has also evaluated it as level 3.The results presented in Table 7 show that core asset development activity is performed at a high maturity level of 4. Product development activity is also performed at a maturity of level 3 to 4, meaning that level 3 has been achieved and level 4 is the next target. The management activity is performed at a lower maturity level of 2. Case study – III confirms that the organization "C" has achieved a CMM-level 3 and that there is a need to improve the management activity to increase the overall maturity level of the organization.

## DISCUSSION OF CASE STUDY – IV

The CMM-level of organization "D" is level 2, i.e., "repeatable", and the software product line process assessment tool have also evaluated it as level 2. Table 8 indicates that core asset development activity is performed at a maturity level of 3. The Product development activity is performed at a higher maturity of level 3 to 4. The management activity is performed at a lower maturity level of 2. The conclusion of the case study – IV pointed out that the organization "D" has achieved a CMM-level 2 and there remains a need to improve the management and core asset development activity to increase the overall maturity level of the organization.

**Table 6. Results of Process Assessment of Case Study - II**

| Activity | Result | Linguistic Output | CMM Level |
|---|---|---|---|
| Core Asset Development Assessment | 37.5 | Managed | 4 |
| Product Development Process Assessment | 37.5 | Managed | 4 |
| Management Process Assessment | 44.67 | Managed To Optimizing | 4 to 5 |
| Software Product Line Process Assessment | 46.11 | Optimizing | 5 |

**Table 7. Results of Process Assessment of Case Study - III**

| Activity | Result | Linguistic Output | CMM Level |
|---|---|---|---|
| Core Asset Development Assessment | 37.5 | Managed | 4 |
| Product Development Process Assessment | 34.84 | Defined to Managed | 3 to 4 |
| Management Process Assessment | 17.5 | Repeatable | 2 |
| Software Product Line Process Assessment | 27.07 | Defined | 3 |

**Table 8. Results of Process Assessment of Case Study - IV**

| Activity | Result | Linguistic Output | CMM Level |
|---|---|---|---|
| Core Asset Development Assessment | 25.65 | Defined | 3 |
| Product Development Process Assessment | 34.84 | Defined to Managed | 3 to 4 |
| Management Process Assessment | 17.5 | Repeatable | 2 |
| Software Product Line Process Assessment | 17.5 | Repeatable | 2 |

## LESSON LEARNED

The case studies presented in this work lead to a number of lessons learned. One of the objectives of this research was to investigate the impact of already achieved CMM level on software product line process. Table 9 shows the comparison of process assessment carried out by using SPLPAT and already achieved CMM level by the organizations under study. We observed from case studies that an organization at a higher CMM level has a better tendency to carry out software product line activities in more effective way. This leads to the conclusion that the higher the

CMM level of an organization, the higher is the software product line process maturity.

The other objective of this study was to investigate how effectively imprecision and uncertainty present in software process data is handled by using fuzzy logic in software product line process assessment. Table 10 illustrates the comparison of the assessment carried out by using statistical average method and fuzzy logic evaluation. The results of both the approaches are compared with already achieved CMM levels of the organizations under study. The lesson learned from this investigation supports the concept that fuzzy

### Table 9. Comparisons of SPLPAT Process Assessment Results with CMM-Level

| Organization | CMM-Level (Already Achieved) | Observed CMM-Level (By SPLPAT) |
|---|---|---|
| "A" | 2 (Repeatable) | 2 (Repeatable) |
| "B" | 5 (Optimizing) | 5 (Optimizing) |
| "C" | 3 (Defined) | 3 (Defined) |
| "D" | 2 (Repeatable) | 2 (Repeatable) |

### Table 10. Comparison of Fuzzy Calculation and Statistical Average Method

| Rule-Input # | Organization & Case Study # | | | |
|---|---|---|---|---|
| | 'A' | 'B' | 'C' | 'D' |
| | Case Study -1 | Case Study -2 | Case Study -3 | Case Study -4 |
| 1 | 35 | 40 | 32.5 | 40 |
| 2 | 40 | 40 | 27.5 | 30 |
| 3 | 25 | 15 | 30 | 35 |
| 4 | 35 | 30 | 37.5 | 30 |
| 5 | 25 | 50 | 40 | 20 |
| 6 | 40 | 15 | 37.5 | 40 |
| 7 | 10 | 15 | 32.5 | 35 |
| 8 | 5 | 30 | 30 | 35 |
| 9 | 50 | 50 | 35 | 30 |
| 10 | 45 | 40 | 37.5 | 30 |
| 11 | 30 | 50 | 32.5 | 25 |
| 12 | 10 | 40 | 35 | 20 |
| 13 | 15 | 40 | 30 | 30 |
| 14 | 20 | 30 | 35 | 35 |
| 15 | 30 | 40 | 32.5 | 35 |
| 16 | 35 | 45 | 30 | 35 |
| 17 | 7 | 25 | 37.5 | 35 |
| Statistical Average | 26.88 (Level 3) | 35 (Level 4) | 33.67 (Level 3 to 4) | 32.23 (Level 3 to 4) |
| Fuzzy (SPLPAT) Calculation | 17.5 (Level 2) | 46.11 (Level 5) | 27.07 (Level 3) | 17.5 (Level 2) |
| Actual CMM | Level 2 | Level 5 | Level 3 | Level 2 |

logic use in software product line process assessment handles imprecision and uncertainty in a much better way, compared to other standard methods, and it provides more reliable assessment.

## CONCLUSION

In this research paper we have tried to identify key process activities in the form of rules for software product line development and management. The developed rules cover several possible aspects that require attention by the technical and organizational management in order to adopt successfully the software product line approach. A software product line process assessment framework is developed to assist an organization to evaluate the current maturity level of the process. On the basis of the developed framework, a tool, software product line process assessment tool (SPLPAT), has been developed to evaluate the current process maturity level within an organization, and direct comparisons are made to CMM. Following are the conclusions and lessons learned during this work:

- The software process maturity assessment determined by using CMM and SPLPAT are the same, a fact that shows the extent of reliability of the proposed approach.

- The lesson learned from case studies presented in this work is that organizations at a higher CMM level have a greater tendency to execute software product line process more effectively.

- Fuzzy logic provides an appropriate approach to handle the uncertainty and imprecision present in the software process variables.

- Software product line rules provide a basis for process maturity level assessment of software product line as they identify key process activities.

- Software product line process assessment tool presented in this work can be used to evaluate the current process status of software product line process within an organization.

**Acknowledgement.** We are thankful to the Institute of Information Technology, National Research Council Canada, for the permission to use the NRC-FuzzyJ Toolkit for research purposes. We are thankful to all those organizations that provided us valuable software product line process data for our research. We are also thankful to those anonymous reviewers and senior editor of this paper for providing us with valuable suggestions and recommendations to improve the paper.

# AUTHORS

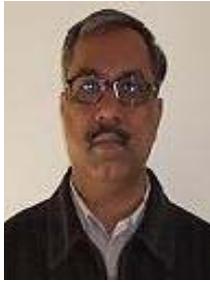

**Faheem Ahmed** is a PhD candidate at University of Western Ontario, Canada, where he received his Masters degree in Electrical Engineering with emphasis in Software Engineering. He received his M.Sc degree in Electronics from Quaid-e-Azam University, Islamabad, Pakistan. Before joining Western as graduate student he has been working in the software industry for 10 years. During his professional career he has been actively involved in requirements engineering, design, development and testing of software products. His current research interests are software engineering, software product line process modeling and process assessment, CASE tools, fuzzy logic, object-oriented design and programming languages. He is a student member of IEEE.

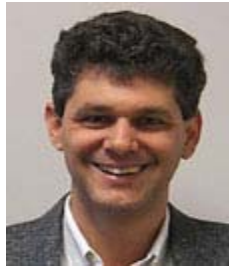

**Dr. Capretz** has over 20 years of experience in the software engineering field as a practitioner, manager and educator. Before joining the University of Western Ontario, in Canada, he worked at both technical and managerial levels, taught and did research on the engineering of software in Brazil, Argentina, England and Japan since 1981. He was the Director of Informatics and Coordinator of the computer science program in two universities (UMC and COC) in the State of Sao Paulo/Brazil. He has authored and co-authored over 50 peer-reviewed research papers on software engineering in leading international journals and conference proceedings, and co-authored the book, *Object-Oriented Software: Design and Maintenance*, published by World Scientific. His current research interests are software engineering (SE), human factors in SE, software product lines, and software engineering education. Dr. Capretz received his Ph.D. from the University of Newcastle upon Tyne (U.K.), M.Sc. from the National Institute for Space Research (INPE-Brazil), and B.Sc. from UNICAMP (Brazil). He is a senior member of IEEE, and a MBTI Qualified Practitioner.